%% file: paper.tex
\documentclass[submission,copyright,creativecommons]{eptcs}

\providecommand{\event}{FOPARA 2017} % Name of the event you are submitting to

\title{Towards Practical, Precise and Parametric \\Energy Analysis of IT Controlled Systems}
\def\titlerunning{Towards Practical, Precise and Parametric energy Analysis of IT Controlled Systems}

\author{
Bernard van Gastel \quad \quad Marko van Eekelen
\institute{Faculty of Management, Science and Technology, \\
Open University of the Netherlands,
Heerlen, The Netherlands}
\institute{Institute for Computing and Information Sciences, \\
Radboud University, Nijmegen, The Netherlands}
\email{\{Bernard.vanGastel,Marko.vanEekelen\}@ou.nl}
}
\def\authorrunning{B. van Gastel and M. van Eekelen}

\input{common.tex}

\default[black]

\usepackage{hyperref}
\hypersetup{colorlinks=true,citecolor=blue,linkcolor=blue,urlcolor=blue}
\usepackage{pgf}
\usepackage{tikz}
\usetikzlibrary{automata}
\usetikzlibrary{arrows}
\usetikzlibrary{positioning}
\usetikzlibrary{fit}

\dutch

\input{definitions.tex}

\begin{document}
\maketitle

\input{abstract}

\input{introduction}

\input{ecaintroduction}

\input{extralanguage}

\input{applications}

\input{relatedwork}
\input{conclusions}

\bibliographystyle{eptcs}
\bibliography{common}

\end{document}

%% file: common.tex
\usepackage{xcolor}
\definecolor{grey}{rgb}{0.6,0.6,0.6}
\definecolor{almostblack}{rgb}{0.3,0.3,0.3}

\definecolor{bitpowderaccent}{HTML}{0077bb}
\definecolor{radbouduniversityaccent}{HTML}{be311a}
\definecolor{openuniversityaccent}{HTML}{e2001a}
\definecolor{dkgreen}{rgb}{0,0.6,0}

\definecolor{mdred}{HTML}{F44336}
\definecolor{mdpink}{HTML}{E91E63}
\definecolor{mdpurple}{HTML}{9C27B0}
\definecolor{mddeeppurple}{HTML}{673AB7}
\definecolor{mdindigo}{HTML}{3F51B5}
\definecolor{mdblue}{HTML}{2196F3}
\definecolor{mdlightblue}{HTML}{03A9F4}
\definecolor{mdcyan}{HTML}{00BCD4}
\definecolor{mdteal}{HTML}{009688}
\definecolor{mdgreen}{HTML}{4CAF50}
\definecolor{mdlightgreen}{HTML}{8BC34A}
\definecolor{mdlime}{HTML}{CDDC39}
\definecolor{mdyellow}{HTML}{FFEB3B}
\definecolor{mdamber}{HTML}{FFC107}
\definecolor{mdorange}{HTML}{FF9800}
\definecolor{mddeeporange}{HTML}{FF5722}
\definecolor{mdbrown}{HTML}{795548}
\definecolor{mdbluegrey}{HTML}{607D8B}
\definecolor{mddeepgreen}{HTML}{388E3C} % actually 700 from "Green"

\definecolor{mdaubergine}{HTML}{C51162} % actually A700 from pink

\definecolor{printred}{cmyk}{0, .9, 1, .2}
\definecolor{printblue}{cmyk}{1, .6, 0, 0}
\definecolor{printdeeppurple}{cmyk}{.65, 1, 0, 0}
\definecolor{printdeeporange}{cmyk}{0, .6, 1, 0}
\definecolor{printzand}{cmyk}{0, .35, 1, 0}
\definecolor{printgreen}{cmyk}{0.9, 0, 1, .4}
\definecolor{printaubergine}{cmyk}{0, 1, .5, .2}
\definecolor{printteal}{cmyk}{0.8, 0, .4, .1}
\definecolor{grey9}{cmyk}{0, 0, 0, 0.9}
\definecolor{grey8}{cmyk}{0, 0, 0, 0.8}
\definecolor{grey7}{cmyk}{0, 0, 0, 0.7}
\definecolor{grey6}{cmyk}{0, 0, 0, 0.6}
\definecolor{grey5}{cmyk}{0, 0, 0, 0.5}
\definecolor{grey4}{cmyk}{0, 0, 0, 0.4}
\definecolor{grey3}{cmyk}{0, 0, 0, 0.3}
\definecolor{grey2}{cmyk}{0, 0, 0, 0.2}
\definecolor{grey1}{cmyk}{0, 0, 0, 0.1}

\usepackage{import}

\newcommand{\edit}[1]{}
\newcommand{\todo}[1]{}

\newcommand{\dutchqed}{\qed}
\newcommand{\dutchfill}{}
\newcommand{\dutch}{
	\providecommand{\setparsizes}[3]{\setlength{\parskip}{##2}\setlength{\parindent}{##1}\renewcommand{\dutchqed}{\qed}}
	\renewcommand{\dutchfill}{\\ \vspace{-\baselineskip}}
	\renewcommand{\dutchqed}{\hfill \qed\dutchfill}
	\setparsizes{0em}{.625\baselineskip plus .125\baselineskip minus .125\baselineskip}{2em plus 1fil}
}

\usepackage{listings}
\lstdefinelanguage{acl2}{
morekeywords={defun,defund,defcong,defthm,defthmd,t,nil,declare,xargs,measure,ignore,defspec,local,defstructure,defconst,defstobj,defabsstobj},
keywordsprefix=:,
otherkeywords={defun-sk,rule-classes,in-theory,begin-book,include-book,in-package,mutual-recursion,instance-of-defspec,make-event},
sensitive=false,
alsodigit=-,
showstringspaces=false,
breaklines=true,
breakatwhitespace=true,
morecomment=[l];,
morecomment=[s]{\#|}{|\#},
morestring=[b]",
stringstyle=\color{accent}}

\lstdefinelanguage{eca}{
morekeywords={for,while,do,repeat,if,then,else,function,begin,end},
otherkeywords={},
sensitive=false,
alsodigit=-,
showstringspaces=false,
breaklines=true,
breakatwhitespace=true,
morecomment=[l]//,
morecomment=[s]{/*}{*/},
morestring=[b]",
stringstyle=\color{accent}}

\lstdefinelanguage{C++}{
morekeywords={for,while,break,do,if,then,else,inline,typedef,enum,struct,void,unsigned,bool,short,int,long,float,double},
otherkeywords={},
sensitive=false,
alsodigit=-,
showstringspaces=false,
breaklines=true,
breakatwhitespace=true,
morecomment=[l]//,
morecomment=[s]{/*}{*/},
morestring=[b]",
stringstyle=\color{accent}}

\lstdefinelanguage{Promela}{
morekeywords={inline,typedef,atomic,bool,assert,chan,if,fi,do,od,else,break,struct,int},
otherkeywords={},
sensitive=false,
alsodigit=-,
showstringspaces=false,
breaklines=true,
breakatwhitespace=true,
morecomment=[l]//,
morecomment=[s]{/*}{*/},
morestring=[b]",
stringstyle=\color{accent}}

\lstdefinelanguage{PVS}{%
    alsoletter={ABCDEFGHIJKLMNOPQRSTUVWXYZabcdefghijklmnopqrstuvwxyz_0123456789},%
    alsoletter={!\#\$\%\&*+-./<=>?@\\^|~:'},%
morekeywords={AND,CODATATYPE,ENDIF,JUDGEMENT,RECURSIVE,ANDTHEN,COINDUCTIVE,ENDTABLE,LAMBDA,SUBLEMMA,ARRAY,COND,EXISTS,LAW,SUBTYPES,AS,CONJECTURE,EXPORTING,LEMMA,SUBTYPE_OF,ASSUMING,    CONTAINING,FACT,LET,TABLE,ASSUMPTION,CONVERSION,FALSE,LIBRARY,THEN,AUTO_REWRITE,CONVERSION+,FORALL,MACRO,THEOREM,AUTO_REWRITE+,CONVERSION-,FORMULA,MEASURE,THEORY,AUTO_REWRITE-CORECURSIVE,FROM,NONEMPTY_TYPE,TRUE,AXIOM,COROLLARY,FUNCTION,NOT,TYPE,BEGIN,DATATYPE,HAS_TYPE,O,TYPE+,BUT,ELSE,IF,OBLIGATION,VAR,BY,ELSIF,IFF,OF,WHEN,CASES,END,IMPLIES,OR,WHERE,CHALLENGE,ENDASSUMING,IMPORTING,ORELSE,WITH,CLAIM,ENDCASES,IN,POSTULATE,XOR,CLOSURE,ENDCOND,INDUCTIVE,PROPOSITION,PRED},%
    sensitive=true,%
    morecomment=[l]{\%},%
    morecomment=[n]{\{-}{-\}},%
    morestring=[b]",%
    emptylines=1,%
    stringstyle=,%
    showstringspaces=false,%
    keepspaces=true,%
    breaklines=false,%
    texcl=true,%
    escapeinside={(\#}{\#)},%
    literate=   {<}{{$<$}}1%
                {>}{{$>$}}1%
                {=}{{$=$}}1%
                {~}{{$\sim$}}1%
                {<=}{{$\leq$}}1%
                {>=}{{$\geq$}}1%
                {/=}{{$\neq$}}1%
                {++}{{$+\!\!\!+$}}2%
                {->}{{$\rightarrow$}}2%
                {<-}{{$\leftarrow$}}2%
                {=>}{{$\Rightarrow$}}2%
                {<=>}{{$\Leftrightarrow$}}2%
                {[]}{{$[\,\,]$}}2%
                {FORALL}{{$\forall$}}1%
                {IMPLIES}{{$\Rightarrow$}}2%
                {EXISTS}{{$\exists$}}1%
                {NOT\ }{{$\neg$}}2%
                {\ AND}{{$\wedge$}}2%
                {\ OR}{{$\vee$}}2%
                { IFF }{{$\Leftrightarrow$}}1%
                {LAMBDA}{{$\lambda$}}1%
                {intersection}{{$\cap$}}1%
                {union}{{$\cup$}}1%
                {emptyset}{{$\emptyset$}}1%
}

\newcommand\defaultinside[0]{\leftskip8mm\rightskip8mm}
\newcommand\dconcrete[1]{\text{\normalfont\ttfamily\color{accent}#1}\xspace} % concrete domain
\newcommand\dcode[1]{\dconcrete{#1}} % code snippets
\newcommand\dfunc[1]{\dconcrete{#1()}} % function
\newcommand\dabstract[1]{\text{\normalfont\sffamily\halfbfseries\color{accent}#1}\xspace} % abstract domain
\newcommand\dtype[1]{\dabstract{#1}} % types (int, double, etc)
\newcommand\dclass[1]{\dabstract{#1}} % class
\newcommand\dlang[1]{\dabstract{#1}} % language feature
\newcommand\term[1]{{\sffamily\color{accent}#1}\xspace} % term

\newcommand{\field}[1]{\ensuremath{\mathrm{#1}}\xspace}

\newcommand{\name}{\ensuremath{\operatorname{name}}\xspace}

\newcommand{\union}[0]{\,\cup\,}

\providecommand{\qed}{\hfill\ensuremath{\square}}

\definecolor{tuplecolor}{gray}{0.5}
\newcommand{\tuplestart}[0]{{\color{tuplecolor}\langle}}
\newcommand{\tupleend}[0]{{\color{tuplecolor}\rangle}}
\newcommand{\tuplesep}[0]{{\color{tuplecolor},\ }}

\newcommand{\pair}[2]{\tuplestart #1\tuplesep #2 \tupleend}
\newcommand{\triple}[3]{\tuplestart {#1}\tuplesep { #2} \tuplesep {#3} \tupleend }
\newcommand{\qple}[4]{\tuplestart {#1}\tuplesep { #2} \tuplesep {#3} \tuplesep {#4} \tupleend }
\newcommand{\qtple}[5]{\tuplestart {#1}\tuplesep { #2} \tuplesep {#3 }\tuplesep {#4} \tuplesep {#5} \tupleend }

\newcommand\defaultpackages{%
        \usepackage{xspace}
        \usepackage{fancyvrb} % better verbatim
	\usepackage{graphicx} %
	\usepackage{amsmath} %
	\usepackage{amsfonts} %
	\usepackage{amssymb} %
	\usepackage{subfig} %
	\usepackage{wasysym} %
	\usepackage{icomma} % , in numbers does not generate a space
	\usepackage[plain,vlined,norelsize]{algorithm2e} % for the algorithm, optional option: ruled
	\usepackage{bussproofs} % for the proof trees
	\usepackage{floatpag} % ability to set \thisfloatpagestyle{empty} for figures
	\usepackage[dutch,english]{babel}\selectlanguage{english}
	\usepackage{multicol}
	\usepackage{wrapfig}
	\usepackage[activate={true,nocompatibility},final,tracking=true,factor=1100,stretch=10,shrink=10]{microtype}
	\usepackage[htt]{hyphenat} % Enable hythenation of TT text:
	\usepackage[overload]{textcase} % working \MakeTextUppercase
	\let\labelindent\relax % disable labeindent in IEEE, as per http://tex.stackexchange.com/questions/170772/command-labelindent-already-defined
	\usepackage{enumitem} % extra options for \begin{description}
	\usepackage{setspace}
	\usepackage{tabularx,colortbl} % tabular environment that can set width; to color rows and tables use extra package 'colortbl'
	\usepackage[all,defaultlines=3]{nowidow} % less widows, especially with align environment
}

\newcommand\defaultpackagesloadedlast{%
        \usepackage{syntax} % for the BNF grammar, needed here because of a conflict with komascript 3.20
        \newlength{\defaultgrammarindent}
        \setlength{\defaultgrammarindent}{10em}
        \setlength{\grammarindent}{\defaultgrammarindent}
        \DeclareTextCommandDefault{\textunderscore}{\leavevmode \kern.06em\vbox{\hrule width .3em}}
}

\newcommand\defaulttheorems{%
	\newtheorem{theorem}{\sffamily\bfseries\color{accent} Theorem} %
	\newtheorem{runningexample}{\sffamily\bfseries\color{accent} Running Example} %
	\newtheorem{example}{\sffamily\bfseries\color{accent} Example} %
	\newtheorem{proof}{\sffamily\bfseries\color{accent} Proof} %
	\newtheorem{exercise}{\sffamily\bfseries\color{accent} Exercise} %
}

\newcommand\defaultsetup{%
	\setlength{\emergencystretch}{2pt} % less overfull hboxes
	\raggedbottom % fix weird behaviour, text below sections can be bottom-aligned, leaving a lot of space between the header and text
	\captionsetup{aboveskip=1mm} % same spacing for all captions
	\AtBeginDocument{
            \setlength{\abovedisplayskip}{1em}
            \setlength{\belowdisplayskip}{1em}
            \setlength{\abovedisplayshortskip}{1em}
            \setlength{\belowdisplayshortskip}{1em}
        }
	\floatpagestyle{empty}% empty page style _only_ for this page
}

\newenvironment{accentrule}[1][]{}{}
\newenvironment{accentruletable}[1][]{}{}
\newlength{\accentruleextra}
\colorlet{accent}{almostblack}
\colorlet{ACCENT}{almostblack} % for titles with \MakeTextUppercase

\makeatletter
\def\provideenvironment{\@star@or@long\provide@environment}
\def\provide@environment#1{%
        \@ifundefined{#1}%
                {\def\reserved@a{\newenvironment{#1}}}%
                {\def\reserved@a{\renewenvironment{dummy@environ}}}%
        \reserved@a
}
\def\dummy@environ{}
\makeatother

\newcommand\defaultstyle[1]{ %
	\colorlet{accent}{#1} %
	\setlength{\accentruleextra}{2mm} %0.001em plus 0em minus 0em} % 0.75em
	\newcommand{\accentrulestart}[1]{{\color{accent}\hrule height 0.25mm\rule{0.25mm}{1.5mm}\hfill\rule{0.25mm}{1.5mm}\vspace{-1.5mm}}\ifthenelse{\equal{##1}{}}{}{\vspace{##1}}} %
	\newcommand{\accentruleend}[1]{\ifthenelse{\equal{##1}{}}{}{\vspace{##1}}\vspace{-1.5mm}{\color{accent}\rule{0.25mm}{1.5mm}\hfill\rule{0.25mm}{1.5mm}\hrule height 0.25mm}} %
	
	\newlength{\parskipsave}
	\newlength{\accentrulethisskip}
	\renewenvironment{accentrule}[1][0mm]{%
		\setlength{\parskipsave}{\parskip}%
		\setlength{\parskip}{0mm}%
		\accentrulestart{}%
		\par%
		\setlength{\accentrulethisskip}{##1}%
		\vspace{\accentrulethisskip}%
	}{%
		\vspace{\accentrulethisskip}%
		\par%
		\accentruleend{}%
		\setlength{\parskip}{\parskipsave}%
	}
				
}

\newcommand\defaultstart[1]{%
	\defaultpackages %
	\defaulttheorems %
	\defaultsetup %
	\defaultstyle{#1} %
}

\newcommand\default[1][almostblack]{%
	\defaultstart{#1}
	\defaultpackagesloadedlast %
}

\newcommand\captionwith[2]{
	\let\originalfigurename\figurename
	\renewcommand\figurename{#1}
	\caption{#2}
	\let\figurename\originalfigurename
}

\newcommand\halfbfseries{}

\newcommand\fancylabel{

\usepackage[most]{tcolorbox}% http://ctan.org/pkg/tcolorbox

\newlength{\accentlabelframethinkness}
\setlength{\accentlabelframethinkness}{0.125mm}
\newlength{\accentlabelradius}
\setlength{\accentlabelradius}{1mm}
\newcommand{\accentstrut}{\vphantom{[y(}}
\newcommand{\accentlabelfont}{\normalfont\sffamily} % with cheap printers, do \bfseries here (premium print)
\newcommand{\accentlabelframefont}{\normalfont\sffamily}

\newcommand{\accentboxhelper}[5]{\begin{tcolorbox}[coltext=white,colframe=##5,colback=##5,boxrule=\accentlabelframethinkness,width=##2+2\accentlabelframethinkness,height=##3+5pt+2\accentlabelframethinkness,arc=\accentlabelradius,left=0mm,right=0mm,top=0mm,bottom=0mm,fontupper=\accentlabelfont,nobeforeafter,baseline=##4+2pt+\accentlabelframethinkness]\centering##1\end{tcolorbox}}

\newcommand{\accentframehelper}[5]{\begin{tcolorbox}[coltext=##5,colframe=##5,colback=white,boxrule=\accentlabelframethinkness,width=##2+2\accentlabelframethinkness,height=##3+5pt+2\accentlabelframethinkness,arc=\accentlabelradius,left=0mm,right=0mm,top=0mm,bottom=0mm,fontupper=\accentlabelframefont,nobeforeafter,baseline=##4+2pt+\accentlabelframethinkness]\centering##1\end{tcolorbox}}

\newlength{\accentboxwidth} % temporary
\newlength{\accentboxheight} % temporary

\newcommand{\accentlabel}[3][]{%
\settowidth{\accentboxwidth}{\accentlabelfont~~##3\accentstrut~~}%
\ifthenelse{\equal{##1}{}}{}{\setlength{\accentboxwidth}{##1}}%
\settoheight{\accentboxheight}{\accentlabelfont ##3\accentstrut}% \strut needed to normalise height, ##1 to incorporate font changes
\accentboxhelper{##3\accentstrut}{\accentboxwidth}{1.2\accentboxheight}{.23\accentboxheight}{##2}% 1.2\baselineskip}%
}

\newcommand{\accentlabelframe}[3][]{%
\settowidth{\accentboxwidth}{\accentlabelframefont~~##3\accentstrut~~}%
\ifthenelse{\equal{##1}{}}{}{\setlength{\accentboxwidth}{##1}}%
\settoheight{\accentboxheight}{\accentlabelframefont ##3\accentstrut}% \strut needed to normalise height, ##1 to incorporate font changes
\accentframehelper{##3\accentstrut}{\accentboxwidth}{1.2\accentboxheight}{.23\accentboxheight}{##2}% 1.2\baselineskip}%
}

\newcommand{\accenttag}[3][]{\accentlabel[##1]{##2}{\footnotesize\bfseries\MakeUppercase{##3}}}
\newcommand{\accenttagframe}[3][]{\accentlabelframe[##1]{##2}{\footnotesize\bfseries\MakeUppercase{##3}}}
}

\newcommand\fancycolors{%
        \colorlet{header}{accent}
        \renewcommand{\syntleft}{$\tuplestart$\normalfont\sffamily\itshape\color{accent}} % for BNF grammars
        \renewcommand{\syntright}{$\tupleend$}
        \renewcommand{\litleft}{\thinspace`\color{accent}\bgroup\ulitleft} % for BNF grammars
        \renewcommand{\litright}{\ulitright\egroup\color{black}'\thinspace}
        \renewcommand{\KwSty}[1]{{\color{accent}\textbf{##1}}} % style algorithm package
        \renewcommand{\CommentSty}[1]{{\color{accent}\texttt{##1}}}  % style algorithm package
        \renewcommand\emph[1]{\textcolor{accent}{\sffamily\bfseries ##1}}
        
        \setlist[description]{leftmargin=8mm,font={\color{accent}}} % \normalfont
        \newcommand{\itemizelabel}{\bfseries---}
        \setlist[itemize]{leftmargin=8mm,label={\color{accent}\itemizelabel}}
        \setlist[enumerate]{leftmargin=8mm,label={\accentlabelframe[6mm]{accent}{\arabic*}}}
        
        	\let\originalfootnoterule\footnoterule
	\renewcommand\footnoterule{{\color{accent}\originalfootnoterule}}
	
}

\newcommand\fancytoc{
\ifcsname chapterpagestyle\endcsname%
    \newcommand*{\SavedOriginaladdchaptertocentry}{}
    \let\SavedOriginaladdchaptertocentry\addchaptertocentry
    \renewcommand*{\addchaptertocentry}[2]{%
      \ifstr{##1}{}{% entry without number
        \SavedOriginaladdchaptertocentry{}{\vspace{-2em} \\ \normalfont\sffamily ##2}%    
      }{% entry with number
        \SavedOriginaladdchaptertocentry{}{\textcolor{accent}{\chaptername~##1} \\ \normalfont\sffamily ##2}%
      }%
    }%
\fi
}

\newcommand\fancybib{
\usepackage[autostyle]{csquotes} % needed for biblatex
\usepackage[%
    style=alphabetic, % numeric-comp
    sorting=myshorthand,
    url=true,%
    backend=biber,% bibtex or biber
    abbreviate=false,%
    maxbibnames=9,%
    defernumbers=true, %
    backref=true, %
    uniquename=false, %
    isbn=false 
]{biblatex} % Load the package with some options.
\DeclareSourcemap{ % adapted from http://tex.stackexchange.com/questions/76534/biblatex-print-isbn-only-if-doi-is-not-defined
  \maps[datatype=bibtex]{
     \map{
        \step[fieldsource=doi,final]
        \step[fieldset=url,null]
        }
      }
}
\DeclareSortingScheme{myshorthand}{%
    \sort{\field{presort}}
    \sort[final]{\field{sortkey}}
    \sort{%
        \field{shorthand}
        \name{sortname}\name{author}\name{editor}\name{translator}
        \field{sorttitle}\field{title}}
    \sort{\field{sortyear}\field{year}}
    \sort{\field{sorttitle}\field{title}}
    \sort{%
        \field[padside=left,padwidth=4,padchar=0]{volume}
        \literal{0000}}}

\appto{\bibsetup}{\sloppy}

\DeclareCaseLangs{british,american,canadian,english,australian,newzealand,USenglish,UKenglish}
\DeclareFieldFormat{titlecase}{\MakeSentenceCase*{##1}}
\newrobustcmd{\MakeTitleCase}[1]{%
  \ifthenelse{\ifcurrentfield{booktitle}\OR\ifcurrentfield{booksubtitle}%
    \OR\ifcurrentfield{maintitle}\OR\ifcurrentfield{mainsubtitle}%
    \OR\ifcurrentfield{journaltitle}\OR\ifcurrentfield{journalsubtitle}%
    \OR\ifcurrentfield{issuetitle}\OR\ifcurrentfield{issuesubtitle}%
    \OR\ifentrytype{book}\OR\ifentrytype{mvbook}\OR\ifentrytype{bookinbook}%
    \OR\ifentrytype{booklet}\OR\ifentrytype{suppbook}%
    \OR\ifentrytype{collection}\OR\ifentrytype{mvcollection}%
    \OR\ifentrytype{suppcollection}\OR\ifentrytype{manual}%
    \OR\ifentrytype{periodical}\OR\ifentrytype{suppperiodical}%
    \OR\ifentrytype{proceedings}\OR\ifentrytype{mvproceedings}%
    \OR\ifentrytype{reference}\OR\ifentrytype{mvreference}%
    \OR\ifentrytype{report}\OR\ifentrytype{thesis}}
    {##1}
    {\MakeSentenceCase*{##1}}}

\DeclareFieldFormat[article,inbook,incollection,inproceedings,patent,thesis,unpublished]{title}{\mkbibquote{\emph{##1}}\isdot}
\DeclareFieldFormat[article]{journaltitle}{\textit{##1}}
\DeclareFieldFormat[inbook,incollection,inproceedings]{booktitle}{\textit{##1}}

\DeclareBibliographyCategory{author}
\DeclareBibliographyCategory{peerreviewed}
\DeclareBibliographyCategory{nonpeerreviewed}
\DeclareBibliographyCategory{media}
\newcommand\authorcite[2]{
	\addtocategory{##1}{##2}
	\addtocategory{author}{##2}
	\nocite{##2}
	}

\defbibenvironment{numbered}
  {\list{}
    {\setlength{\leftmargin}{\bibhang}%
    \setlength{\itemindent}{-\leftmargin}%
    \setlength{\itemsep}{\bibitemsep}%
    \setlength{\parsep}{\bibparsep}}}
  {\endlist}
  {\item}

\providecommand{\noopsort}[1]{}

}

%% file: definitions.tex
\newcommand{\rcomposition}[0]{>\!\!>\!\!>}

\newcommand{\access}{::}
\newcommand{\accessfield}{.}

\newcommand{\AxiomB}[1]{\AxiomC{}\UnaryInfC{{#1}}}

\newcommand{\subst}[3]{ {{#1}[{#3} \gets {#2}]} } % mapsto
\newcommand{\scopeho}[3]{ \overline{[{#3} \mapsto {#2}, {#1}]} } % mapsto

\newcommand{\langvariable}{x}

\newcommand{\langcomponent}{C}
\newcommand{\langfunction}{f}

\newcommand{\langassign}[2]{{#1}\text{\dlang{~:=~}}{#2}}
\newcommand{\langexpression}{e}

\newcommand{\langfunctiondefinition}[3]{\text{\dlang{function}} \ #1\text{\dlang{(}}{#2}\text{\dlang{)}} \ \text{\dlang{begin}} \ {#3} \ \text{\dlang{end}}}

\newcommand{\langclassfunctionname}{{fun-name}}
\newcommand{\langclasscomponentname}{{component}}
\newcommand{\langclassfunctiondef}{{fun-def}}

\newcommand{\langclassexpression}{{expr}}
\newcommand{\langclassvariable}{{var}}
\newcommand{\langclassconstant}{{const}}

\newcommand{\langclassstatement}{{stmt}}

\newcommand{\langbinop}{{bin-op}}

\newcommand{\expressionreduction}{~{\xrightarrow{e}}~}

\newcommand{\componentstates}{{\Gamma}}

\newcommand{\eca}{\term{ECA}}

\newcommand{\simplescalar}{\term{SimpleScalar}}

\newcommand{\PState}{\operatorname{PState}}
\newcommand{\GState}{\operatorname{GState}}
\newcommand{\CState}{\operatorname{CState}}
\newcommand{\pstate}{\mathit{ps}}
\newcommand{\gstate}{\mathit{gs}}
\newcommand{\cstate}{\mathit{cs}}

%% file: abstract.tex
\begin{abstract}
Energy consumption analysis of IT-controlled systems can play a major role in minimising the overall energy consumption of such IT systems, during the development phase, or for optimisation in the field. Recently,  a precise energy analysis was developed, with the property of being parametric in the hardware. In principle, this creates the opportunity to analyse which is the best software implementation for given hardware, or the other way around: choose the best hardware for a given algorithm.  

The precise analysis was introduced for a very limited language: \term{ECA}. In this paper, several important steps are taken towards practical energy analysis. The \term{ECA} language is extended with common programming language features. The application domain is further explored, and threats to the validity are identified and discussed.  Altogether, this constitutes an important step towards analysing energy consumption of IT-controlled systems in practice.

%Energy analysis consists of a hybrid approach in analysing both hardware and software together, to derive energy consumptions when executing the software on the hardware. This can be used during the development phase, or for optimisation. One can e.g. choose the best software implementation for given hardware, or the other way around: choose the best hardware for a given algorithm.
%
%Previously in \cite{DBLP:conf/fopara/GastelKE15} a program transformation was proposed, that could derive energy consumption of programs written in \term{ECA}. However, the language supported by the analysis was limited. In this article we extend the language support of the analysis with recursion, global variables and types.
\end{abstract}

%% file: introduction.tex
%%%%%%%%%%%%%%%%%%%%%
\section{Introduction}
\label{sect:intro}
%%%%%%%%%%%%%%%%%%%%%A
Energy analysis of IT systems is an important emerging field. Its focus is on analysing the software that controls the IT system using models of the components of the system under analysis.  Components can vary from small components such as a sensor in the Internet of Things to large subsystems as present in self-driving cars.

As traditionally many savings did occur on the hardware side of a computer, energy consumption is almost a blind spot when developing software. Each next hardware generation consumed less energy to perform the same amount of work. However, recently this development has lost its pace. At the same time, it becomes more and more clear that software has a huge impact on the behaviour and the properties of devices it runs on. A recent example of software influencing the working of a device is the Volkswagen scandal. The car manufacturer used software to detect if the car was being tested.
%and if so, made the diesel motor exhaust less toxic gases
If this was found to be the case, the diesel motor was programmed to operate in such a way that it exhausted less toxic gases and fumes. In \cite{Oldenkamp2016121} it is calculated that 44,000 years of human life are lost in Europe because of the fraud, which lasted at least six years. Another example is fridges from Panasonic, which could detect if a test was going on and suppressed energy intensive defrost cycles during this test. These are negative examples, but they do make clear that the software is in control of the device and its (energy) behaviour. 
% http://www.nu.nl/internet/4260652/browser-opera-komt-met-energiebesparende-functie.html

Although the software is evidently in control of the devices, there is almost no time dedicated in most computer science curricula to the energy efficiency of software. 
This is peculiar since energy is of vital importance to the modern (software) industry. For years, data centres have been located at places where the energy is cheap, and since the rise of the smartphone more software engineers recognise that to get good user reviews, their software should not rapidly deplete the battery charge of the user's phone. Due to this lack of educational attention to energy-aware programming, most aspiring programmers never learn to produce energy efficient code. Software engineers have trouble assessing how much energy will be consumed by their software on a target device, especially when the software is run on a multitude of different systems.
With the advent of the internet of things, where software is increasingly embedded in our daily life, the \emph{software industry should become aware of their energy footprint}, and methods must be developed to assist in reducing this footprint.

% energie, kleine wijzigingen -> groot effect
Furthermore, the combination of many individual negative effects can also affect our society at large. Although this effect is less direct, it is no less essential. If devices that are present in large quantities in our society all exhibit the same negative behaviour, such as incurring needlessly a too high energy consumption, they can impact public utilities and our economy and will consume the finite resources of Earth even faster. Governments increasingly recognise this societal effect, as indicated by the new laws in the European Union issuing ecodesign requirements for many kinds of devices.
%aiming among others to make them more energy efficient.
One of the aims of these requirements is to make devices more energy efficient.
Examples of product categories with ecodesign requirements include vacuum cleaners, electrical motors, lighting, heaters, cooking appliances, televisions and coffee machines. Even requirements leading to relative small improvements in energy efficiency can yield large results at scale, %even for devices one typically would not expect to be able to save electricity significantly.
even in the case of devices of which one would expect no significant electricity savings to be possible. 

Modern devices and appliances are controlled by software, which makes analysing the energy consumption challenging of these devices, as the behaviour of its software is difficult to predict. To analyse the consumption of hardware, the software controlling the hardware needs to be analysed together with the hardware.

\paragraph{Our approach} To this end, we proposed in \cite{DBLP:conf/fopara/GastelKE15} a hybrid approach, joining energy behaviour models of the hardware with the energy-aware semantics of software and a program transformation. The interface between hardware and software is made explicit and has to be well defined, allowing for exchanging of hardware or software components. Using this parametric approach, multiple implementations can be analysed.
Such an approach can be used on design level or for optimisation.
One can e.g.\ choose the best software implementation for given hardware, or the other way around: choose the best hardware for a given algorithm.

The described approach derives energy consumption functions. These energy functions signify the exact energy behaviour when the software is executing and controlling the modelled hardware. Hardware is modelled as a finite state machine, with both the states and transitions labelled with energy consumption. The programs that can be analysed are written in the software language \term{ECA}, which is an imperative language inspired by \term{C} and \term{Java}. Currently, only a limited set of program constructs is supported.
%Key to analysing larger systems is compositionality.

The most important contributions of this article are:
\begin{itemize}[topsep=0pt]
  \item extended support for common language features in the software to be analysed: adding types, data structures, global variables, and recursion;
  \item description of application domain of the \eca energy analysis method;
  \item identified threats to the validity of proposed approach and a discussion of how to deal with these threats.
\end{itemize}

\paragraph{Overview} Section \ref{sect:eca-intro} introduces the \term{ECA} energy analysis. In section \ref{sect:extend} extensions of the \term{ECA} language are defined including the new derivation rules that are needed for the analysis. Section \ref{sect:applic} explores the application domain of \term{ECA}.  The validity of the results of the analysis is discussed in section \ref{sect:energy-discussion}. Finally, we conclude with related work, future work and conclusions in sections \ref{sect:rel-fut} and \ref{sect:energy-conclusion}.

%% file: ecaintroduction.tex
%%%%%%%%%%%%%%%%%%%%%
\section{Introduction to energy analysis with \term{ECA}}
\label{sect:eca-intro}
%%%%%%%%%%%%%%%%%%%%%

Energy analysis combines hardware modelling with the energy-aware semantics of software. To this end the language \term{ECA} is specified, on which our analysis is targeted. Based on this, a semantics of this language can be defined, which includes energy consumption. Using this energy-aware semantics, a program transformation is given. This transformation generates an executable model, using a parametric function. If both a concrete input and one or multiple hardware models are specified, the parametric function will result in the energy consumption occurring when running the software on the given hardware.

To illustrate this process, we start with the hardware modelling and continue with describing for some program constructs the semantic rules, the program transformations and the effect on hardware. This is an introductory overview. For further details, the reader is referred to \cite{DBLP:conf/fopara/GastelKE15,BernardPhD,fopara14-kersten}.

The hardware conceptually consists of a \emph{component state} and a set of \emph{component functions} which operate on the component state.
We use finite state models to model these hardware components, with the transitions constituting function calls on the hardware.
Energy usage is expressed by labelling both the vertices and edges with energy consumption, which can be in any unit.
Labels on vertices constitute time-bound energy consumption, i.e.\ power draw. Edges are labelled with the consumption of a certain amount of energy, not time-bound but corresponding to the transition. Depending on your needs, you can model energy consumption in one way or the other.  
Besides the ones above described, there are no additional requirements. We use the power draw function $\phi$ which translates a component state to a power draw. The result of this function is used to calculate energy consumption for the time spent in a specific state.

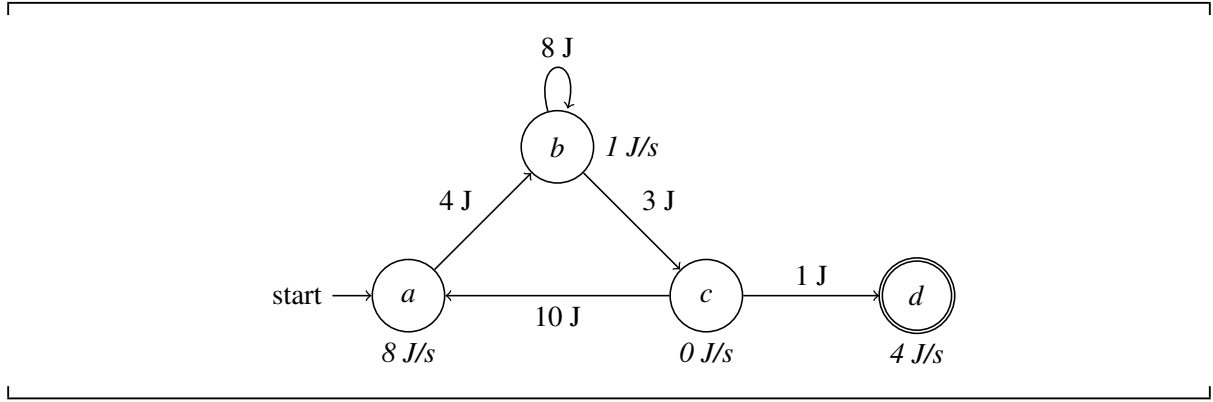
\begin{figure}[!h]
\begin{accentrule}
  \begin{center}
\begin{tikzpicture}[->,auto,node distance=2.8cm,semithick]
  \node[initial,state,label={below:\textit{8 J/s}}]    (A)                    {$a$};
  \node[state,label={right:\textit{1 J/s}}]            (B) [above right of=A] {$b$};
  \node[state,label={below:\textit{0 J/s}}]            (C) [below right of=B] {$c$};
  \node[state,accepting,label={below:\textit{4 J/s}}]  (D) [right of=C]       {$d$};
  \path (A) edge              node {4 J} (B)
        (C) edge              node {10 J} (A)
        (B) edge [loop above] node {8 J} (B)
            edge              node {3 J} (C)
        (C) edge              node {1 J} (D);
\end{tikzpicture}
  \end{center}
  \vspace{-0.75em}
\end{accentrule}
  \vspace{0.5em}
  \caption{Hardware model with the energy consumption expressed in Joule.}
\label{ecagrammarrepeat}
\end{figure}

All transitions in a component model are explicit in the \term{ECA} source code. We use the notation $C \text{\dlang{\access}} f$ to refer to a function \dfunc{f} operating on a component \dclass{C}. Multiple different hardware components can be used simaltaniously from the same program, the components are differentiated by a unique name (substituted in the rules for \dclass{C}). Besides this addition, the \term{ECA} language is a fairly default imperative language sporting functions, a single signed integer type, conditionals, looping constructs and expressions that can be used as statements.

Next, we move on to the semantics, for now without energy added. Besides the function environment $\Delta$ and the program state $\sigma$, we have the (hardware) component states $\Gamma$. This makes the effect on the hardware explicit.
The effect of the component function $C \text{\dlang{\access}} f$ is split into two: the effect function $\delta_{\,C \text{\dlang{\access}} f}$ and a function $\operatorname{rv}_{C \text{\dlang{\access}} f}$ calculating the return value of the component call. Both are working on the component state retrieved from $\Gamma$ (by $\Gamma(C)$).
The straightforward semantic definition of the component function \emph{sCmp}, not taking energy into account, is given below.
$$
\AxiomC{$ \Delta \vdash \triple{\langexpression_1}{\sigma}{\componentstates} \expressionreduction \triple{a}{\sigma'}{\componentstates'} $}
\RightLabel{(sCmp)}
\UnaryInfC{$ \Delta, \langcomponent\text{\dlang{\access}}\langfunction\!=\!(\delta_{\;C \text{\dlang{\access}} f}, \operatorname{rv}_{C \text{\dlang{\access}} f}) \vdash  \triple{ \langcomponent\text{\dlang{\access}}\langfunction\text{\dlang{(}}\langexpression_1\text{\dlang{)}} }{\sigma}{\componentstates} 
              \expressionreduction \triple{\operatorname{rv}_{C \text{\dlang{\access}} f} (\componentstates'(\langcomponent), a)}{\sigma'}{\componentstates'[\langcomponent \gets \delta_{\;C \text{\dlang{\access}} f}(\componentstates'(\langcomponent), a)]} $}
\DisplayProof
$$

Adding to it, the energy cost of a component function call consists of including the time taken to execute this function and the explicit energy cost attributed to this call resulting in rule \emph{esCmp}.
$$
\AxiomC{$ \Delta; \Phi \vdash \triple{\langexpression_1}{\sigma}{\componentstates} \expressionreduction \qple{a}{\sigma'}{\componentstates'}{E'} $}
\RightLabel{(esCmp)}
\UnaryInfC{$ \Delta, \langcomponent\text{\dlang{\access}}\langfunction\!=\!(\mathellipsis, \mathellipsis, t\!_f, E\!_f); \Phi \vdash \\ \triple{ \langcomponent\text{\dlang{\access}}\langfunction\text{\dlang{(}}\langexpression_1\text{\dlang{)}} }{\sigma}{\componentstates} 
              \expressionreduction \qple{\mathellipsis}{\mathellipsis}{\mathellipsis}{E' + \Phi(\componentstates') \cdot t\!_f + E\!_f} $}
\DisplayProof
$$
The approach works by transforming these semantic rules into higher order expressions. When executed on a concrete program state $\PState$ and component state $\CState$, this expression yields the energy consumption (and new states and possible value of an \term{ECA} expression). Expressions from the \term{ECA} language are transformed into rules that result in a tuple of three elements: a value function $V$, a state update function $\Sigma$ (for both program state and the hardware state), and an energy consumption function $E$. Statements from the \term{ECA} language are only transformed into the latter two. 

These compositional expressions are composed with higher order combinators. One of these combinators is the composition operator $\rcomposition$, which first applies the left-hand side, and on the resulting state applies the right-hand side. Another is the $\overline{+}$ operator, which is a higher order addition operator: when executed, it calculates the energy consumptions of the two operands based on the input states and adds them together. 
To explain the component call rule \emph{btCmp} and function call rule \emph{btCall}, we need an operator for higher order scoping. This operator creates a new program environment but retains the component state. It can even update the component state given a $\Sigma$ function, which is needed because this $\Sigma$ needs to be evaluated using the original program state. The definition is as follows:
\begin{align*}
&    \scopeho{\Sigma}{V}{x}: \operatorname{Var} \times (\PState \times \CState \rightarrow \operatorname{Value}) & \\
& \qquad \times (\PState \times \CState \rightarrow \PState \times \CState) & \\
& \qquad \rightarrow (\PState \times \CState \rightarrow \PState \times \CState) & \\
& \scopeho{\Sigma}{V}{x}(\pstate, \cstate)  =  ([x \mapsto V(\pstate, \cstate)], \cstate') \text{ where } (\_, \cstate') = \Sigma(\pstate, \cstate) &
\end{align*}

We also need an additional operator $\operatorname{split}$, because the program state is isolated, but the component state is not. The $\operatorname{split}$ function forks the evaluation into two state update functions and joins the results together. The first argument defines the resulting program state; the second defines the resulting component state.
\begin{align*}
\RightLabel{(btCmp)}
  \AxiomC{$\Delta \vdash \langexpression : \triple{V_{ex}}{\Sigma_{ex}}{\mathellipsis}$}
\UnaryInfC{$
\begin{array}{rl}
  \Delta, & \hspace{-0.5em} \langcomponent\text{\dlang{\access}}\langfunction = (\langvariable\!_{f}, V\!\!_{f}, \Sigma\!_{f}, \mathellipsis, \mathellipsis) \vdash \\
  & \hspace{-0.5em} \langcomponent\text{\dlang{\access}}\langfunction\text{\dlang{(}}\langexpression\text{\dlang{)}} : \triple{\scopeho{\Sigma_{ex}}{V_{ex}}{x\!_{f}} \rcomposition V\!\!_{f}}{\operatorname{split}(\Sigma_{ex}, \scopeho{\Sigma_{ex}}{V_{ex}}{x\!_{f}} \rcomposition \Sigma\!_{f})}{\mathellipsis}
\end{array}
$} 
 \DisplayProof
\end{align*}

The environment $\Delta$ is extended for each component function $\langcomponent \access f$ with two elements: an energy judgment $E\!_{f}$ and a run-time $t_{f}$. Time independent energy usage can be encoded into this $E\!_{f}$ function. For functions defined in the language, the derived energy judgement is inserted into the environment.
%There is no need for these functions for an explicit run-time as this is part of the derived energy judgement.
Using the patterns described above the component function call is expressed as:
$$
%\fontsize{8}{10}\selectfont
\RightLabel{(etCmp)}
\AxiomC{$\Delta \vdash \langexpression : \triple{V_{ex}}{\Sigma_{ex}}{E_{ex}}$}
\UnaryInfC{
$
\begin{array}{rl}
  \Delta, & \hspace{-0.5em} \langcomponent\text{\dlang{\access}}\langfunction = (\langvariable\!_{f}, \mathellipsis, \mathellipsis, E\!_{f}, t\!_{f}) \vdash \\
  & \hspace{-0.5em} \langcomponent\text{\dlang{\access}}\langfunction\text{\dlang{(}}\langexpression\text{\dlang{)}} : \triple{\mathellipsis}{\mathellipsis}{E_{ex}\ \overline{+}\ (\Sigma_{ex} \rcomposition (\operatorname{td}^{ec}(t\!_{f})\ \overline{+}\ E\!_{f}))}
\end{array}
$
}
\DisplayProof
$$

This concludes the short introduction to energy analysis with \term{ECA}. For a more thorough coverage, see~\cite{DBLP:conf/fopara/GastelKE15,BernardPhD,fopara14-kersten}.

%% file: extralanguage.tex
%%%%%%%%%%%%%%%%%%%%%
\section{Increasing the expressivity of  \term{ECA} }
\label{sect:extend}
%%%%%%%%%%%%%%%%%%%%%

To bridge the gap between practical programming languages and \term{ECA}, several extensions to \term{ECA} are introduced in this section: adding data structures and types, global variables and recursion.

\subsection{Adding data structures and types}
\begin{figure}[!b]
\begin{accentrule}
\setlength{\grammarindent}{7.5em}
\begin{grammar}
\defaultinside
<program> ::=  <struct-def> <program> | <\langclassfunctiondef> <program> | <type> <\langclassvariable> `=' <\langclassexpression> | $\epsilon$

<struct-def> ::= `struct' <struct-name> `begin' <struct-fields> `end'

<struct-fields> ::= <type> <field-name> `;' <struct-fields> | $\epsilon$

<type> ::= `void' | `bool' | `int' | `float' | <struct-name>

<\langclassfunctiondef> ::= <type> <\langclassfunctionname> `(' [<fun-args>] `)' `begin' <\langclassexpression> `end'

<fun-args> ::= <type> <name> `,' <fun-args> | <type> <name>

<\langbinop> ::= `+' | `-' | `*' | `>' | `>=' | `==' | `!=' | `<=' | `<' | `and' | `or'

<\langclassexpression> ::= <\langclassconstant> | <\langclassvariable> | <\langclassexpression> <\langbinop> <\langclassexpression>
\alt <struct-name> `(' <args> `)'
\alt <\langclassexpression> `\accessfield' <field-name>
\alt <type> <\langclassvariable> `=' <\langclassexpression>
\alt <\langclassvariable> `=' <\langclassexpression>
  \alt <\langclasscomponentname> `\access' <\langclassfunctionname> `(' [<args>] `)'
  \alt <\langclassfunctionname> `(' [<args>] `)'
\alt <\langclassstatement> `,' <\langclassexpression>

  <args> ::= <\langclassexpression> `,' <args> | <\langclassexpression>

  <\langclassstatement> ::= `skip' | <\langclassstatement> `;' <\langclassstatement> | <\langclassexpression>
  \alt `if' <\langclassexpression> `then' <\langclassstatement> [`else' <\langclassstatement>] `end'
  \alt `repeat' <\langclassexpression> `begin' <\langclassstatement> `end'
  \alt `while' <\langclassexpression> `begin' <\langclassstatement> `end'
%\alt <\langclassfunctiondef> <\langclassstatement>

\end{grammar}
\setlength{\grammarindent}{\defaultgrammarindent}
\end{accentrule}
  \vspace{0.5em}
  \captionwith{\lstlistingname}{Extended BNF grammar for the \eca language, with types and data structures added, as well as one construct in \synt{program} for global variable support (see next section).}
\label{ecagrammarrepeat}
\end{figure}

The only supported type in the \eca language was a signed integer. There were no explicit Booleans, floating point numbers or data structures. To add those, types of variables need to be supported. We need multiple modifications for this change: modifying the grammar, and adding type distinction to both the semantic environments and program transformations.
  
We consider variables to be passed \emph{by-value}. Functions can have side effects on the components and, as introduced in section~\ref{sect:globalvariables}, on the global variables. Functions are statically scoped.
\emph{Recursion is now supported}, and the changes needed to the semantic rules and program transformations are discussed in section~\ref{sect:recursion}.

The extended BNF grammar for the \eca language is defined in listing~\ref{ecagrammarrepeat}. We presume there is a way to differentiate between identifiers that represent variables \synt{\langclassvariable}, function names \synt{\langclassfunctionname}, components \synt{\langclasscomponentname}, and constants \synt{\langclassconstant}.

Functions on components can now have a variable number of arguments and optionally return a value (if not, the type \dtype{void} should be used). A constructor for data structures is included, with a syntax like a function call with as name the type of the data structure), and a construct to access fields of a data structure (with the~\dcode{.} operator). A type checking phase is now needed to detect typing errors, like using a data structure as a condition in the \dcode{if} construct. Only correctly typed programs are considered.
The language retains an explicit construct for operations on hardware components (e.g.\ memory, storage or network devices).
The notation $C \text{\dlang{\access}} f$ refers to a function \dfunc{f} operating on a component \dclass{C}. 
This allows us to reason about components in a straightforward manner.

A typical (predictive recursive descent) parser of this extended language is in the $\operatorname{LL}(2)$ class of parsers, with a small second pass. This second pass is needed, to avoid a possible infinite lookahead that is needed to differentiate between expressions and statements. During the first phase expressions and statements are combined into one construct. The small post-processing step differentiates between the two. In this way, the language can still be efficiently parsed in a simple manner.

Next are the adjustments to the semantic rules. Because a type checking phase was added, no typing error can occur when applying the semantic rules. Although the meaning differs, the syntax of the rules remains largely the same. Likewise, we adjust the program transformation rules. To support the new grammar rules, we add additional rules to the existing body of rules. Below is the rule for field access on a variable listed.
$$
\RightLabel{(esField)}
\AxiomC{$ \Delta; \Phi \vdash \triple{\langexpression}{\sigma}{\componentstates} \expressionreduction \qple{v}{\sigma'}{\componentstates'}{E'} $}
\UnaryInfC{$\Delta; \Phi \vdash \triple{\langexpression \dcode{\accessfield} a}{\sigma}{\componentstates} \expressionreduction \qple{v(a)}{\sigma'}{\componentstates'}{E' + \Phi(\componentstates') \cdot t_\text{fieldaccess}}$}
\DisplayProof
$$

%%%%%%%%%%%%%%%%%%%%%
\subsection{Global variables}
\label{sect:globalvariables}
%%%%%%%%%%%%%%%%%%%%%

Control software often works with global variables. To support analysis of this control software we consequently need support in our \term{ECA} language for global variables. Hardware components are already handled as global state.
To also support global variables, we need to introduce an additional global program state environment in addition to the local program state environment as it is currently used.

In the semantics, an additional global program state $G$ is added to all tuples in every rule.
Lookups are first performed in the local scope, the already existing $\sigma$. Although for scoping a layered program state can be preferred, or one based on indirections, we use a different approach. Because of special handling of the, by definition global, hardware components, the global program state is handled in the same manner as the global hardware component states.
%If not found in the local, they are tried in the global scope.
We add rules in the semantics for global variable definitions. The assignment rule is split depending if you assign a global or local variable. The variable loop is adjusted to first look in the local scope and if nothing is found, continue in the global scope. We start with introducing the global assignment rule.
$$
\AxiomC{$ \Delta; \Phi \vdash \qple{\langexpression_1}{\sigma}{G}{\componentstates} \expressionreduction \qtple{n}{\sigma'}{G'}{\componentstates'}{E'}$}
%\AxiomC{$ \Delta^{es}; \Phi \vdash  \resfasgn(\componentstates',\cy') = (\componentstates'',\cy'') $}
\RightLabel{(esGlobAssign)}
\UnaryInfC{$ \Delta; \Phi \vdash \qple{\langassign{\langvariable}{\langexpression_1}}{\sigma}{G}{\componentstates} 	\expressionreduction
                \qtple{n}{\sigma'}{\subst{G'}{n}{\langvariable}}{\componentstates'}{E' + \Phi(\componentstates') \cdot t_\text{assign}}
$}
\DisplayProof
$$

Next, we adjust the variable lookup rule, with a $\union$ defined on two program environments. This $\union$ creates one environment, according to the scoping rules. It a variable is defined in both, the left-hand argument to the $\union$ is normative.
$$
\RightLabel{(esVar)}
\AxiomB{$\Delta; \Phi \vdash \triple{\langvariable}{\sigma}{G}{\componentstates} \expressionreduction \qtple{(\sigma \union G)(\langvariable)}{\sigma}{G}{\componentstates}{\Phi(\componentstates) \cdot t_\text{var}}$}
\DisplayProof
$$

For the program transformation rules, more extensive changes are needed. We extend the (global) component states $\CState$ with a global program state of type $\PState$. For clarity of presentation, we introduce the type $\GState$ (global state) which is a combination of $\PState \times \CState$. In this way, only the higher order combinators need to be changed, as the rules themselves remain intact. Most changes are only to the signature of those combinators.
As a result, these combinators retain their compositional properties.
As an example, the higher order scoping rule of section~\ref{sect:eca-intro} is redefined below.
\begin{align*}
&    \scopeho{\Sigma}{V}{x}: \operatorname{Var} \times (\PState \times \GState \rightarrow \operatorname{Value}) & \\
& \qquad \times (\PState \times \GState \rightarrow \PState \times \GState) & \\
& \qquad \rightarrow (\PState \times \GState \rightarrow \PState \times \GState) & \\
& \scopeho{\Sigma}{V}{x}(\pstate, \gstate)  =  ([x \mapsto V(\pstate, \gstate)], \gstate') \text{ where } (\_, \gstate') = \Sigma(\pstate, \gstate) &
\end{align*}

The $\operatorname{lookup}$ function that the variable lookup rule depends on is redefined as follows:
\begin{align*}
    & \operatorname{lookup}_{x}: \PState \times \GState \rightarrow \operatorname{Value} & \\
    & \operatorname{lookup}_{x}(\pstate, \gstate) = \pstate(x) & \text{ if $x$ exists in local program scope $\pstate$}  \\
    & \operatorname{lookup}_{x}(\pstate, \gstate) = \operatorname{Variables}(\gstate)(x) & \text{ if $x$ exists in the global variables part of $\gstate$} 
\end{align*}

%%%%%%%%%%%%%%%%%%%%%
\subsection{Adding recursion}
\label{sect:recursion}
%%%%%%%%%%%%%%%%%%%%%
We can define the function call in a similar way as the component call, which was introduced in section \ref{sect:eca-intro}. However, to support recursion, a special $\operatorname{subst}$ higher order function is introduced to unfold the function definition once, just before it is executed on a concrete environment. The $\operatorname{subst}$ is defined as follows, with $\PState$ signifying a program state, $\GState$ signifying the global state (extended in section~\ref{sect:globalvariables} to be both the component states and the global variables) and $T$ a type variable (depending on whether a state update or a value function is substituted):
\begin{align*}
& \operatorname{subst}: (\PState \times \GState \rightarrow T) & \\
& \qquad \times (\PState \times \GState \rightarrow T) & \\
& \qquad \rightarrow (\PState \times \GState \rightarrow T) & \\
& \operatorname{subst}(T, R)(\pstate, \gstate)  =  (\subst{T}{\operatorname{subst}(T, R)}{R})(\pstate, \gstate) &
\end{align*}
A recursive call is represented by the abstract higher-order function $\operatorname{rec}$, which is a placeholder for applying substitution on. There are multiple variants, depending on the resulting type, with the $\operatorname{rec}_V$ one for a resulting value function, and the $\operatorname{rec}_\Sigma$ one for a resulting state update function.
\begin{align*}
& \operatorname{rec}_V: \PState \times \GState \rightarrow \operatorname{Value} & \\
& \operatorname{rec}_\Sigma: \PState \times \GState \rightarrow \PState \times \GState &
\end{align*}
If there is a function body $B$ computing a $\operatorname{Value}$ with for example $\operatorname{rec}_V$ in it, the value of the recursive function can be computed by executing $\operatorname{subst}(B,\operatorname{rec}_V)$. As long as the original function terminates on the given input environment, this analysis will terminate on the same input. This is the essential difference from the \emph{btCmp} rule, as can be seen in the definition of \emph{btCall} below:
\begin{align*}
\small
\AxiomC{$\Delta, f\!=\!(\langvariable\!_{f}, V\!\!_{f}, \Sigma\!_{f}) \vdash \langexpression : \pair{V_{ex}}{\Sigma_{ex}}$}
\RightLabel{(btCall)}
\UnaryInfC{$
\begin{array}{rl}
\Delta^v, f\!=\!(\langvariable\!_{f}, V\!\!_{f}, \Sigma\!_{f}) \vdash \langfunction\text{\dlang{(}}\langexpression\text{\dlang{)}} : \pair{ & \hspace{-1em}\scopeho{\Sigma_{ex}}{V_{ex}}{x\!_{f}} \rcomposition \operatorname{subst}(V\!\!_{f}, \operatorname{rec}_V(f))}{\\
& \hspace{0.5em} \operatorname{split}(\Sigma_{ex}, \scopeho{\Sigma_{ex}}{V_{ex}}{x\!_{f}} \rcomposition \operatorname{subst}(\Sigma\!_{f}, \operatorname{rec}_\Sigma(f)))\ }
\end{array}
$}
\DisplayProof
\end{align*}
For each language function, a definition is placed in $\Delta^v$ using the \emph{btFuncDef} rule. The body of the function is analysed, and recursive calls to the function are replaced with $\operatorname{rec}$ placeholders using the \emph{btRec} rule. To support this, the function definition rule inserts a special definition in the function environment $\Delta^v$, on which the \emph{btRec} rule works. This leads to the following definition of \emph{btFuncDef}, with $P$ the remaining program definition:
\begin{align*}
\small
\AxiomC{$\Delta^v, f\!=\!(\langvariable) \vdash \langexpression : \pair{V_{ex}}{\Sigma_{ex}}$}
\AxiomC{$\Delta^v, f\!=\!(\langvariable, V_{ex}, \Sigma_{ex}) \vdash P : \Sigma_{st}$}
\RightLabel{(btFuncDef)}
\BinaryInfC{$\Delta^v \vdash  \langfunctiondefinition{f}{\langvariable}{\langexpression} \ P : \Sigma_{st}$}
\DisplayProof
\end{align*}
The placeholders are inserted using the \emph{btRec} rule. This rule analyses the expression used as the argument, like the component and function call rules do. The definition is in fact very similar to those definitions:
\begin{align*}
\small
\AxiomC{$\Delta^v, f\!=\!(\langvariable\!_{f}) \vdash \langexpression : \pair{V_{ex}}{\Sigma_{ex}}$}
\RightLabel{(btRec)}
\UnaryInfC{$
\begin{array}{rl}
\Delta^v, & \hspace{-1em} f\!=\!(\langvariable\!_{f}) \vdash \\
& \hspace{-1em} \langfunction\text{\dlang{(}}\langexpression\text{\dlang{)}} : \pair{\scopeho{\Sigma_{ex}}{V_{ex}}{x\!_{f}} \rcomposition \operatorname{rec}_V(f)}{\operatorname{split}(\Sigma_{ex}, \scopeho{\Sigma_{ex}}{V_{ex}}{x\!_{f}} \rcomposition \operatorname{rec}_\Sigma(f))}
\end{array}
$}
\DisplayProof
\end{align*}

%% file: applications.tex
%%%%%%%%%%%%%%%%%%%%%
\section{Exploring the application domain of \term{ECA}}
\label{sect:applic}
%%%%%%%%%%%%%%%%%%%%%
The foreseen application area of the proposed analysis is in predicting the energy consumption of control systems, where software controls peripherals. This includes control systems in factories, cars, aeroplanes, smart-home applications, etc. Examples of hardware components range from heaters to engines, motors and urban lighting.
Depending on the target device energy consumption can be electricity, gas, water, or any other resources where the consumption increases monotonically.
The proposed analysis can predict the energy consumption of multiple algorithms and different hardware configurations. The choice of algorithm or configuration may depend on the expected workload. This makes the proposed technique useful for both programmers and operators.
Below, we discuss the application domain of \term{ECA} in a way which is partly and informally published in the lecture notes for the TACLe PhD summer school in 2016 in Yspertal, Austria~\cite{energysummerschool}. 

%The used hardware models need to be precise. 
The possibility to abstract from the actual hardware specification makes the proposed approach still applicable even when no final hardware component is available for basing the hardware model on, or when such a model is not yet created. We observe that many decisions are based on relative properties between systems. Abstracting hardware models can be used to focus e.g.\ on the relative differences between component methods and component states. 
%Hardware models must satisfy the properties described in Section~\ref{sect:modelling}.

Compared to the Hoare logic in \cite{fopara14-kersten}, many restrictions are not present in \term{ECA}. Foremost, this type system does not have the limitation that state change cannot depend on the argument of a component function nor that the return value of a component function cannot depend on the state of the component. More realistic models can, therefore, be used. This widens the number of applications, as behaviour of hardware can be modelled that previously could not be expressed in the modelling. 

However, there are still certain properties hardware models must satisfy for \term{ECA} to be applicable. Foremost, the models have to be discrete. Energy consumption that gradually increases or decreases over time can therefore not be modelled directly. However, discrete approximations may be used. 
Secondly, every state change has to be the consequence of an explicit application of a component function. So, implicit state changes by hardware components cannot be expressed. 

The quality of the derived energy expressions is directly related to the quality of the used hardware models.
Depending on the goal, it is possible to use multiple models for one and the same hardware component. For instance, if the hardware model is constructed as a worst-case model, this approach will produce worst-case information. Similarly one can use average-case models to derive average case information. 

It can be difficult to obtain detailed hardware models, sometimes for the simple reason that the hardware is yet to be developed and not ready. We expect that, in cases where multiple software implementations are to be compared, the relative consumption information will be sufficient to support design decisions. This allows for constructing abstract models, with not much detail but including the relevant information which is needed to make a proper comparison. However, this abstraction could impact the validity of the results (as a realistic model could yield different results).

A class of applications where the approach described in this article could be useful is a company that produces many variations of the same device. Variations occur based on local requirements, or on regional differences in the electric grid, or on different requirements set by integrators or consumers, or any combination thereof. The \term{ECA} approach allows for quickly designing those variations, and having a clear view on how changes will impact all those variations.

%%%%%%%%%%%%%%%%%%%%%
\section{Validity}
\label{sect:energy-discussion}
%%%%%%%%%%%%%%%%%%%%%

The analysis is sound and complete as can be proven by induction on the syntactic structure of the program in a similar way as in an earlier version of our analysis described in~\cite{techrep}, in which the proof is more complex due to the presence of approximations. An analysis method may be in itself sound and complete but the validity of applying the method in practice can not automatically be inferred from that.

There are several validity constraints to the technique as it is discussed in this article. The quality of the results depends directly on the quality of the component models used. There are severe restrictions on component models, e.g.\ the power draw is assumed to be constant in every state of the component. This is in practice not true for most devices, e.g.\ the power draw can be a function of time. This has to be modelled in an abstract discretised component model. It is to be seen whether with discrete approximations for real world hardware component models can be created with a level of precision that is suited to make accurate energy estimates.
%To correctly derive a bound, component models should be bound on their maximum energy usage by other means. To this end, alternative techniques can potentially be applied, such as model transformations and timed model checking, however this is left for future research. 
It is hard to construct and validation such component models on the right level of abstraction, as there are several real world practical issues.
If basing the component model on specifications from hardware vendors, all kinds of errors in the specification are transferred to the component model.
Production errors in the hardware, and eventually the degradation of hardware, can induce erratic energy consumption behaviour that does not conform to the specification/component model.

Validating a component model with a test setup is hard, as energy is hard to measure. Small differences in energy consumption are hard to measure correctly, and outside conditions like temperature can influence the results greatly. Energy differs significantly from other kinds of resources (e.g.\ memory and time), which are measurable with great precision within a computer by the computer itself. Introducing a standard energy consumption measurement interval might help in making measurements  more uniform.
Validating if the number of states of a component model is the same as the actual number of states of an actual hardware component, is a hard problem by itself. With powerful models, the actual validation process with real hardware might just take too much time forcing the user to settle for a feasible but  not fully validated model.

There is another potential source of not matching the actual energy consumption of a realistic situation. Compiler errors and optimisations can impact the (energy) behaviour of a source program greatly. The compiler has influence on the timing of high-level language constructs. The timing constants used for these language constructs should match the time it takes to execute those language constructs. Such a match could be guaranteed by creating a resource consumption certified compiler in a similar way as was done in the CompCert certified compiler project~\cite{Leroy2009:Compcert}. Of course this would require the availability of energy aware semantics both on the source and the target level. An even more complicating matter may be the complex design of modern processors executing the software. Even relatively small embedded microprocessors have features (register bypass e.g.), which impact the execution timing of statements significantly. Proper documentation of such features may be hard to find since e.g. the inner details of the pipeline of modern CPU's can often not be found in the documentation.

These constraints on modelling hardware components and validity implications should be lifted and further investigated to make the technique discussed applicable to general, real-world problems.
However, depending on the context and the precision needed, the current technique can already be applicable now. If the hardware component is relatively simple, a suited component model can be constructed. Another valid area for the techniques discussed is to give feedback to a prospective programmer, such that during construction of software the developer can optimise the energy consumption for various hardware configurations.

%% file: relatedwork.tex
%%%%%%%%%%%%%%%%%%%%%
\section{Related work and future work}
\label{sect:rel-fut}
%%%%%%%%%%%%%%%%%%%%%

A few options are available to a programmer who wants to write energy-efficient code.
The programmer can look for programming guidelines and design patterns, which in most cases produce more energy-efficient programs, e.g.~\cite{Saxe10,DBLP:conf/sac/BrinkeMBBA13}. Then, he/she might make use of a compiler that optimises for energy-efficiency, e.g.~\cite{Zhurikhin09}. If the programmer is lucky, there is an energy analysis available for the specific platform at hand, such as~\cite{JayaseelanML06} in which the energy consumption of a processor is modelled in \simplescalar.

However, for most platforms, this is not a viable option. In that case, the programmer might use dynamic analysis with a measurement set-up. This, however, is not a trivial task and requires a complex set-up~\cite{Jagroep:2016:SEP:2889160.2889216,seflab}. Moreover, it only yields information for a specific benchmark~\cite{7101364}. Nevertheless, these approaches are always applicable. A programmer might, however, prefer an approach that yields additional insight in a more predictive manner.

In future work, we aim to fully implement this precise analysis to evaluate its suitability for larger systems and further explore practical applicability. We intend to experiment with additional implementations of various derived approximating analyses to evaluate which techniques/approximations work best in which context.

A current limitation of the analysis is that it allows only \emph{one control process} (processor). Actual systems often consist of a network of interacting systems. Therefore, incorporating interacting systems would increase the applicability of the approach. Such systems can be seen as hybrid automata. Theoretical and practical results in modelling hybrid automata~\cite{DBLP:conf/hybrid/AlurCHH92,DBLP:conf/hybrid/AlurMT16} might be a useful starting point for further research.

To make the presentation more concise, it might be useful to use as a subject language a first-order strict-evaluation functional programming language. One can expect that this will alleviates the need to have a separate basic dependent type system which transforms all variables into expressions over input variables. However, expressions would still need to be expressed in terms over input variables. 
To support the analysis of data types, a size analysis of data types might be useful to enable iteration over data structures, e.g.\ using techniques similar to~\cite{DBLP:conf/fopara/ShkaravskaET13,Tamalet2009}. %\cite{LMCS_polyShapelySizeAnalysis2009}

On the language level, the type system is precise. However, it does not take into account optimisations and transformations below the language level. This can be achieved by analysing the software on a lower level, for example, the intermediate representation of a modern compiler, or even the binary level. For increased accuracy, this may certainly be worthwhile.
Another motivation to use such an intermediate representation as the language that is analysed is the ability to support (combinations of) many higher level languages. In this way, programs written in and consisting of multiple languages can be analysed. It can also account for optimisations (such as common subexpression elimination, inlining, statically evaluating expressions), which in general reduce the execution time of the program and therefore impact the time-dependent energy usage (calls with side effects like component function calls are not optimised).

%Another future research direction is to expand the supported language that is analysed. %Currently, recursion is not supported. To add recursion to the language that is analysed, an approach is to use the function signatures to compose a set of cost relations (a special case of recurrence relations). A recurrence solver can then eliminate the recursion in the resulting function signatures.

Compared with \cite{fopara14-kersten}, the \term{ECA} energy consumption in this paper is precise instead of over-approximated. Future research can show if the expressions derived by this type system can be transformed in such a way that also upper bound expressions are derived. To support this, recursion and loops should be transformed in a \term{Cost Relation System} (CRS), a special case of recurrence relations. Solving this \term{CRS}, one can acquire a direct formula expressing the energy consumption of the recursive function or loop. 

Another approach to providing more precise estimates is described in \cite{10.1109/ECRTS.2015.17}. In this approach suitable program inputs are identified through which, by measurement, more precise results can be achieved in combination with an auxiliary energy model taking into account the energy consumption of instructions in relation to each other.

Finally, a systematic approach to constructing component models should be looked into. One can create a model from the specifications given by the vendor. Another way is using model learning~\cite{DBLP:journals/cacm/Vaandrager17} techniques, which create a finite state model from black box testing and measuring. All the states should have a time dependent energy consumption assigned to them, and all the transitions should be assigned incidental energy consumption. Such a model can then be used as a component model.

%% file: conclusions.tex
%%%%%%%%%%%%%%%%%%%%%
\section{Conclusion}
\label{sect:energy-conclusion}
%%%%%%%%%%%%%%%%%%%%%
Energy analysis consists of a hybrid approach in analysing both hardware and software together, to derive energy consumptions when executing the software on the hardware. This can be used during the development phase, or for optimisation. One can e.g.\ choose the best software implementation for given hardware, or the other way around: choose the best hardware for a given algorithm.
The key to analysing larger systems is compositionality.
Many programs encountered in the real world feature language constructs such as global variables and recursion. To analyse these programs, support in the \term{ECA} language and program transformations is required.
This article extends \term{ECA} with language support for multiple types, support for recursive functions and global variables.
Important properties are retained: it remains a composable, precise, and parametric energy analysis.

Furthermore, to gain additional insights in the feasibility of the approach, the article explores
the (im)possibilities of using the approach in practice by discussing validity and applications.

All in all, this constitutes a practical step towards the application of the proposed energy analysis on real world problems.